\documentclass[epj]{webofc}
\usepackage[varg]{txfonts}   % Web of Conferences font
\woctitle{Physics Opportunities at an Electron-Ion Collider}%
\begin{document}
\title{Probing the transversity spin structure of a nucleon in neutrino-production of a charmed meson}

\author{B. Pire\inst{1}\fnsep\thanks{\email{bernard.pire@polytechnique.edu}} \and
        L. Szymanowski\inst{2}\fnsep\thanks{\email{lech.szymanowski@ncbj.gov.pl}} \and
        J. Wagner\inst{2}\fnsep\thanks{\email{jakub.wagner@ncbj.gov.pl}}
        % etc.
}

\institute{Centre de physique th\'eorique,  \'Ecole Polytechnique, CNRS, Universit\'e Paris-Saclay, 91128 Palaiseau, France
\and
           National Centre for Nuclear Research (NCBJ), Warsaw, Poland
 }

\abstract{%
Including   $O(m_c)$ terms in the coefficient functions and/or $O(m_D)$ twist 3 contributions in the heavy meson distribution amplitudes leads to a non-zero transverse amplitude for  exclusive  neutrino production of a $D$ pseudoscalar charmed meson on an unpolarized target. We work in the framework of the collinear QCD approach where chiral-odd transversity generalized parton distributions (GPDs) factorize from  perturbatively calculable coefficient functions.  
}
\maketitle
\section{Introduction}
\label{intro}

The now well established framework of collinear QCD factorization \cite{fact1,fact2,fact3} for exclusive reactions mediated by a highly virtual photon in the generalized Bjorken regime describes hadronic amplitudes using generalized parton distributions (GPDs) which give access to a 3-dimensional analysis \cite{3d} of  the internal structure of hadrons.  
Neutrino production is another way to access (generalized) parton distributions  \cite{weakGPD}. Although neutrino induced cross sections are orders of magnitudes smaller than those for electroproduction and neutrino beams are much more difficult to handle than charged lepton beams, they have  been very important to scrutinize the flavor content of the nucleon and the advent of new generations of neutrino experiments opens new possibilities. In particular, the flavor changing character of the electroweak current allows charmed quark to be produced in processes involving light quark partonic distributions  \cite{PS}. This in turn allows helicity flip hard amplitudes to occur at the $O(\frac{m_c}{Q})$ level where $Q$ is the typical large scale allowing QCD collinear factorization. Such a coefficient function has to be attached to a chiral-odd generalized parton distribution, the elusive transversity GPDs \cite{transGPDdef, transGPDno, transGPDacc}. The transverse character of these GPDs select the transverse polarization of the $W-$boson, which phenomenologically allows a separation of this interesting amplitude through the azimuthal distribution of the final state particles \cite{PS}.  

\section{Kinematics}
\label{sec-1}

For definiteness, we consider the exclusive production of a pseudoscalar $D-$meson through the reaction  (see Fig. 1):
\begin{eqnarray}
\nu_l (k)\,N(p_1) &\to& l^- (k')\,D^+ (p_D)\,N(p_2) \,,
\end{eqnarray}
where N is a proton or a neutron, in the kinematical domain where collinear factorization  leads to a description of the scattering amplitude 
in terms of nucleon GPDs and the $D-$meson distribution amplitude, with the hard subprocesses:
\begin{eqnarray}
W^+ (\varepsilon, q)\,d &\to& D^+ (p_D)\,d\,.
\end{eqnarray}
Our kinematical notations are as follows ($m$ and $M_D$ are the nucleon and $D-$meson masses, $m_c$ will denote the charmed quark mass):
\begin{eqnarray}
&&q=k-k'~~~~~; ~~~~~Q^2 = -q^2~~~~~; ~~~~~\Delta = p_2-p_1 ~~~~~; ~~~~~\Delta^2=t \,;\nonumber\\
&&p_1^\mu=(1+\xi)p^\mu +\frac{1}{2}  \frac{m^2-\Delta_T^2/4}{1+\xi} n^\mu -\frac{\Delta_T^\mu}{2}~~~~;~~~~ p_2^\mu=(1-\xi)p^\mu +\frac{1}{2}  \frac{m^2-\Delta_T^2/4}{1-\xi} n^\mu +\frac{\Delta_T^\mu}{2}\,; \\
&&q^\mu= -2\xi' p^\mu +\frac{Q^2}{4\xi'} n^\mu ~~~~;~~~p_D^\mu=  2(\xi-\xi') p^\mu +\frac{M_D^2-\Delta_T^2}{4(\xi-\xi')}  n^\mu -\Delta_T^\mu \,,\nonumber
\end{eqnarray}
with $p^2 = n^2 = 0$ and $p.n=1$. As in  the double deeply virtual Compton scattering case~\cite{DDVCS}, it is meaningful to introduce two distinct momentum fractions:
 \begin{eqnarray}
\xi = - \frac{(p_2-p_1).n}{2} ~~~~~,~~~~~\xi' = - \frac{q.n}{2} \,.
\end{eqnarray} 
Neglecting the nucleon mass and $\Delta_T$, the approximate values of $\xi$ and $\xi'$ are
\begin{eqnarray}
\xi \approx \frac{Q^2+M_D^2}{4p_1.q-Q^2-M_D^2} ~~~~,~~~  \xi' \approx \frac{Q^2}{4p_1.q-Q^2-M_D^2} \,.
\end{eqnarray}
To unify the description of the scaling amplitude, we define a modified Bjorken variable
$
x_B^D \equiv \frac {Q^2+M_D^2}{2p_1.q} 
$
which allows to express $\xi$ and $\xi'$ in a compact form:
 \begin{eqnarray}
\xi \approx \frac{x_B^D }{2-x_B^D } ~~~~~,~~~~~\xi' \approx \frac{x_B }{2-x_B^D } \,.
\end{eqnarray}
If the meson mass is the relevant large scale (for instance in the limiting case where $Q^2$ vanishes as in the timelike Compton scattering kinematics \cite{TCS}) :
\begin{eqnarray}
Q^2\to0 ~~;~~\xi' \to 0 ~;~ \xi \approx \frac{\tau}{2-\tau} ~~;~~\tau = \frac{M_D^2}{s_{WN}-m^2}\,.
\end{eqnarray}

\section{The transverse amplitude}
\begin{figure}
\centering
\includegraphics[width=10cm,clip]{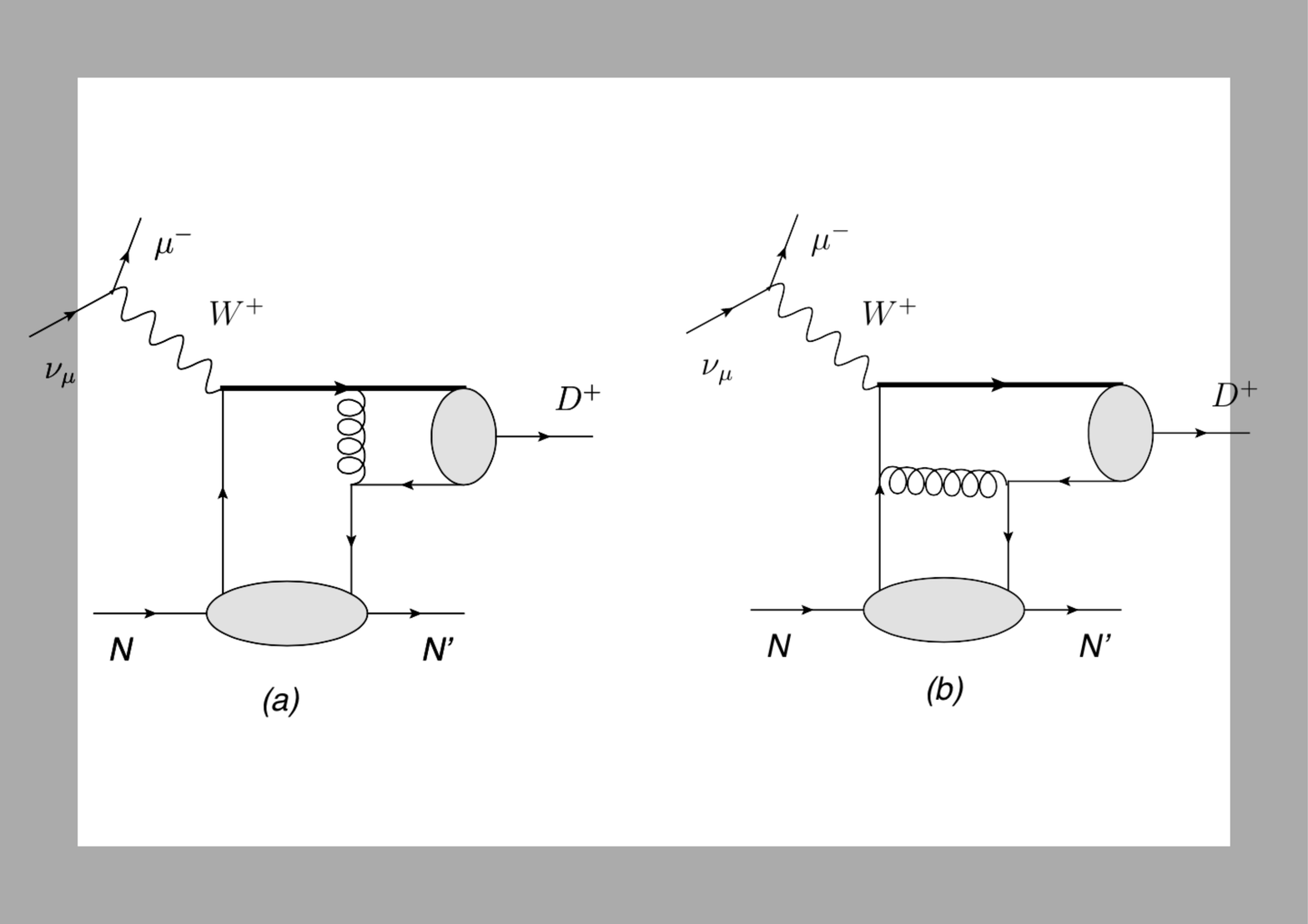}
\caption{Feynman diagrams for the factorized  amplitude for the $ \nu_\mu N \to \mu^-  D^+ N'$ or the $ \nu_\mu N \to \mu^-  D^0 N'$ process involving the quark GPDs; the thick line represents the heavy quark. In the Feynman gauge, diagram (a) involves convolution with both the transversity GPDs and the chiral even ones, whereas diagram (b) involves only chiral even GPDs.}
\label{fig-1}       % Give a unique label
\end{figure}

In the Feynman gauge, the non-vanishing $m_c-$dependent part of the Dirac trace in the hard scattering part depicted in Fig. 1a reads:
\begin{eqnarray}
Tr[\sigma^{pi}\gamma^\nu \hat p_D \gamma^5 \gamma^{\nu'} \frac{m_c}{D_1}\hat \varepsilon (1- \gamma^5)\frac{-g_{\nu \nu'}}{D_2} ]
= \frac{2 (Q^2+M_D^2)}{\xi}  \varepsilon_\mu[ \epsilon^{\mu p i n} + i g^{\mu i}_\perp]   \frac{m_c}{D_1} \frac{1}{D_2} \, ,
 \end{eqnarray} 
 where 
 % the upper (resp. lower) sign  in $\pm$ or $\mp$  stands for neutrino (resp. antineutrino) scattering, 
 $\varepsilon$ is the polarization vector of the $W^\pm$boson (we denote $\hat p = p_\mu \gamma^\mu$ for any vector $p$).
The fermionic trace vanishes for the diagram shown on Fig. 1b thanks to the identity $\gamma^\rho \sigma^{\alpha \beta}\gamma_\rho = 0$. The denominators of the propagators 
read:
\begin{eqnarray}
&&D_1 = k_c^2 - m_c^2 +i\epsilon= \frac{Q^2 }{2 \xi'} (x+\xi-2\xi') - m_c^2 +i\epsilon =  \frac{Q^2+M_D^2 }{2 \xi} (x+\xi) - Q^2   - m_c^2 +i\epsilon \,,\\
 && D_2 = k_g^2 +i\epsilon= \bar z [\bar z m_D^2 + \frac{Q^2 + m_D^2}{2 \xi} (x-\xi) +i\epsilon]\, , \nonumber
 \end{eqnarray}  
 where $k_c$ ($k_g$) is the heavy quark (gluon) momentum. The transverse amplitude is then written  as ($\tau = 1-i2$):
\begin{eqnarray}
T_{T}  = \frac{iC \xi (m_c-2M_D)}{\sqrt 2 (Q^2+M_D^2)}  \bar{N}(p_{2}) \left[  {\mathcal{H}}_{T}^\phi i\sigma^{n\tau} +\tilde {\mathcal{H}}_{T}^\phi \frac{\Delta^{\tau}}{m_N^2}  
+ {\mathcal E}_{T}^\phi \frac{\hat n \Delta ^{\tau}+2\xi  \gamma ^{\tau}}{2m_N} - \tilde {\mathcal E}_{T}^\phi \frac{\gamma ^{\tau}}{m_N}\right] N(p_{1}), 
\label{tramp}
\end{eqnarray}
with $C= \frac{2\pi}{3}C_F \alpha_s V_{dc}$, in terms of  transverse form factors that we define as  :
\begin{eqnarray}
{\cal F }_T^\phi=f_{D}\int \frac{\phi(z)dz}{\bar z}\hspace{-.1cm}\int \frac{F^d_T(x,\xi,t) dx }{(x-\xi+\beta \xi+i\epsilon) (x-\xi +\alpha \bar z+i\epsilon)},
\label{TFF}
 \end{eqnarray} 
where $F^d_T$ is any d-quark transversity GPD, $\alpha = \frac {2 \xi M_D^2}{Q^2+M_D^2}$, $\beta =  \frac {2 (M_D^2-m_c^2)}{Q^2+M_D^2}$ and we shall denote ${\bar {\mathcal{E}}_T^\phi}=\xi{\mathcal{E}}_T^\phi-{\tilde {\mathcal{E}}_T^\phi}$ . In the following, we shall put $\beta$ to $0$.

The prefactor in Eq.(\ref{tramp}) shows the two sources of the transverse amplitude : $m_c$ signals the contribution from the helicity changing part of the heavy quark propagator, while $M_D$ signals the contribution from the twist 3 heavy meson distribution amplitude which we parametrize (omitting the Wilson lines) as :
\begin{eqnarray}
  \langle 0 | \bar c(y)~\gamma^5~d(-y) |D^- (P_D) \rangle  =
  - i  \frac{f_D}{4} M_D \int^1_0 dz~ e^{i(2z-1) P_D.y}~ \phi^{(s)}_D(z)\,, 
         \end{eqnarray}
 and for simplicity we identify $\phi^s$ with the leading twist 2 pseudoscalar charmed meson DA $\phi_D$ defined as:
$$
\langle 0 | \bar c(y)~ \gamma^\mu \gamma^5~d(-y) |D^-(P_D) \rangle  =
 - i  \frac{f_D}{4}  P_D^\mu \int^1_0 dz~e^{i(2z-1) P_D.y}   \phi_D(z)\,. $$

\section{The azimuthal dependence of  neutrinoproduction.}
The dependence of a leptoproduction  cross section on azimuthal angles is a  widely used way to analyze the scattering mechanism. This procedure is helpful as soon as one can define an angle $\varphi$ between a leptonic and a hadronic plane, as for deeply virtual Compton scattering \cite{DGPR} and related processes. In the neutrino case, it reads :
\begin{eqnarray}
\label{cs}
\frac{d^4\sigma(\nu N\to l^- N'D)}{dx_B\, dQ^2\, dt\,  d\varphi}&=&
%-\varepsilon\cos(2\varphi)\sigma_{+-}
\\
 \tilde\Gamma
\Bigl\{ ~\frac{1+ \sqrt{1-\varepsilon^2}}{2} \sigma_{- -}&+&\varepsilon\sigma_{00}+  \sqrt{\varepsilon}(\sqrt{1+\varepsilon}+\sqrt{1-\varepsilon} )(\cos\varphi\
{\rm Re}\sigma_{- 0} + \sin\varphi\
 {\rm Im}\sigma_{- 0} )\ \Bigr\},\nonumber
\end{eqnarray}
with 
\begin{equation}
\tilde \Gamma = \frac{G_F^2}{(2 \pi)^4} \frac{1}{16x_B} \frac{1}{\sqrt{ 1+4x_B^2m_N^2/Q^2}}\frac{1}{(s-m_N^2)^2} \frac{Q^2}{1-\epsilon}\,, \nonumber
\end{equation}
and the ``cross-sections'' $\sigma_{lm}=\epsilon^{* \mu}_l W_{\mu \nu} \epsilon^\nu_m$ are product of  amplitudes for the process $ W(\epsilon_l) N\to D N' $, averaged  (summed) over the initial (final) hadron polarizations.
In the anti-neutrino case,  one gets
a similar expression with $\sigma_{--} \to \sigma_{++}$ , $\sigma_{-0}\to \sigma_{+0}$, $1+ \sqrt{1-\varepsilon^2} \to 1- \sqrt{1-\varepsilon^2}$ and $\sqrt{1+\varepsilon}+\sqrt{1-\varepsilon} \to \sqrt{1+\varepsilon}-\sqrt{1-\varepsilon}$.
We use the standard notations of deep exclusive leptoproduction, namely  $y=p_1.q/p_1.k$ and $\epsilon \simeq 2(1-y)/[1+(1-y)^2]$.  The azimuthal angle $\varphi$ is defined  in the  initial nucleon  rest frame as: 
\begin{equation}
sin ~\varphi = \frac {\vec q \cdot[(\vec q \times \vec p_D) \times (\vec q \times \vec k)]}{|\vec q||\vec q\times \vec p_D||
\vec q\times \vec k|}\,,
\end{equation}
while the final nucleon momentum lies in the $xz$ plane ($\Delta^y = 0$).

The quantity  $\sigma_{- 0} $ is directly related to the observables $<cos~\varphi>$ and $<sin~\varphi>$ through
     \begin{eqnarray}
  <cos ~\varphi>&=&\frac{\int cos ~\varphi ~d\varphi ~d^4\sigma}{\int d\varphi ~d^4\sigma}= \frac{\sqrt \varepsilon (\sqrt{1+\varepsilon}+\sqrt{1-\varepsilon} )\, {\cal R}e \sigma_{- 0}}{2\epsilon \sigma_{0 0} + (1+ \sqrt{1-\varepsilon^2}) \sigma_{--}}  \,, \nonumber \\
   <sin~ \varphi>&=&  \frac{\int sin ~\varphi ~d\varphi ~d^4\sigma}{\int d\varphi ~d^4\sigma}= \frac{\sqrt \varepsilon (\sqrt{1+\varepsilon}+\sqrt{1-\varepsilon} )\, {\cal I}m \sigma_{- 0}}{2\epsilon \sigma_{0 0} + (1+ \sqrt{1-\varepsilon^2}) \sigma_{--}}      \,.
   \end{eqnarray} 
\begin{figure}
\centering
\includegraphics[width=10cm,clip]{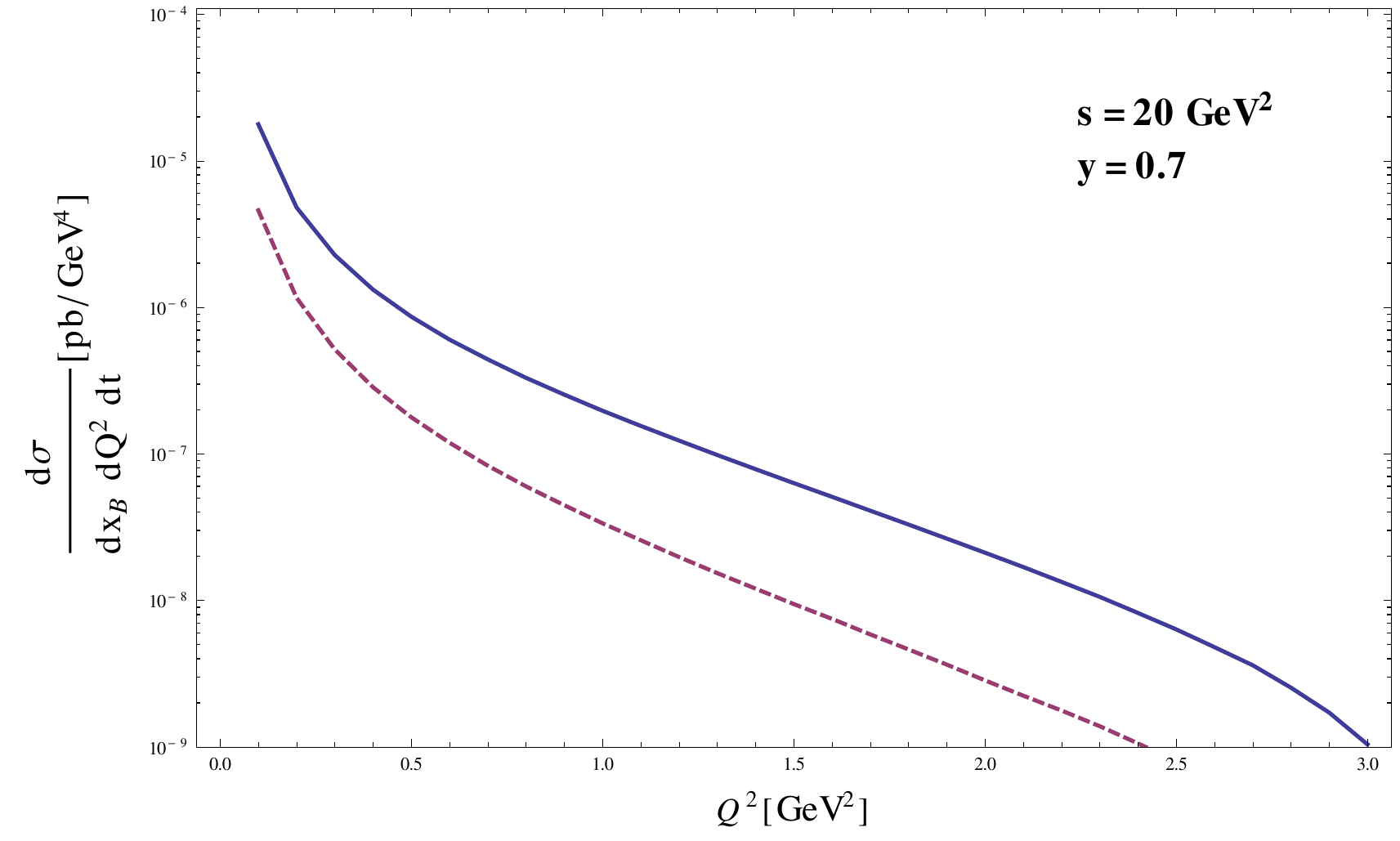}
\caption{Contribution of $\sigma_{ - -}$ to the  differential cross section $\frac{d\sigma}{dx_B dQ^2 d t}$  of neutrino-production of a $D^+$ meson on a proton (dashed line) or a neutron (solid line) at $s = 20$ GeV$^2$ , $y=0.7$ and $t  =t_0$, as a function of $Q^2$.
}
\label{figresult}       % Give a unique label
\end{figure}

Estimating the counting rates and the angular observables defined above is in progress \cite{prog}.   As a first step, we calculate $\sigma_{--} $ which is bilinear in transversity quark GPDs. At zeroth order in $\Delta_T$, $\sigma_{--}$ reads:
\begin{eqnarray}
\sigma_{--} =   \frac{4\xi^2 C^2 (m_c+2M_D)^2}{(Q^2+M_D^2)^2}\biggl\{(1-\xi^2)|{\mathcal{H}_T^\phi}|^2  + \frac{\xi^2}{1-\xi^2} | {\bar {\mathcal{E}}_T^\phi}|^2 -2\xi \mathcal{R}e [ \mathcal{H}_T^\phi {\bar {\mathcal{E}}_T^{\phi *}}]\biggr\} . 
\end{eqnarray}
Using the model of Ref \cite{heavyDA2} for the $D^+$ meson distribution amplitude and the parametrization of the dominant transversity GPD $H_T(x,\xi,t)$ from Ref \cite{ElB} (and neglecting for the time being other chiral-odd GPDs contributions), we compute the contribution to the differential cross section given in Eq.(\ref{cs}) integrated over $\varphi$. The result is shown in Fig. \ref{figresult} as a function of $Q^2$ for $s=20$ GeV$^2$, $y=0.7$ and $t=t_{min}$. Since the process selects the $d-$ quark contribution, the proton and neutron target cases allow to access $H_T^d$ and $H_T^u$ respectively. Although small, the cross-sections are of the same order of magnitude as those for the neutrino production of $\pi$  or $D_s$ mesons estimated in \cite{weakGPD}. This shows that these processes should be measurable in intense neutrino beam facilities.

Let us remind the reader that we allow $Q^2$ to be quite small since the hard scale governing our process is $M_D^2+Q^2$.

\section{Conclusion.}
 Collinear QCD factorization has allowed us  to calculate neutrino production of $D-$mesons in terms of GPDs. Gluon and both  chiral-odd and chiral-even quark GPDs contribute to the amplitude for different polarization states of the $W^\pm$ boson. The azimuthal dependence of the cross section allows to separate different contributions. 
Planned high energy neutrino facilities \cite{NOVA} which have their scientific program oriented toward the understanding of neutrino oscillations  or elusive inert neutrinos may thus allow - without much additional equipment - some important progress in the realm of hadronic physics. 
 
\vspace{.2cm}
This work  was partially supported by the COPIN-IN2P3 Agreement and by the french grant ANR PARTONS (Grant No. ANR-12-MONU-0008-01); L. Sz was partially supported by grant of National Science Center, Poland, No. 2015/17/B/ST2/01838.

\end{document}